\documentclass[final,5p,times,twocolumn]{elsarticle}
\pdfoutput=1
\usepackage{graphicx}
\usepackage{gensymb}

\usepackage{amssymb}

\usepackage{txfonts}

\journal{Nuclear Instruments and Methods A}

\begin{document}

\begin{frontmatter}



\title{Evaluation of two thermal neutron detection units consisting of ZnS/${}^6$LiF scintillating layers with embedded WLS fibers read out with a SiPM}



\author[A]{J.-B. Mosset\corref{cor1}}\ead{jean-baptiste.mosset@a3.epfl.ch}
\author[A]{A. Stoykov}
\author[A]{U. Greuter}
\author[A]{M. Hildebrandt}
\author[A]{N. Schlumpf}
\author[A,B]{H. Van Swygenhoven}
\address[A]{Paul Scherrer Institute, CH-5232 Villigen PSI, Switzerland}
\address[B]{Ecole Polytechnique F\'{e}d\'{e}rale de Lausanne, CH-1015 Lausanne, Switzerland}
\cortext[cor1]{Corresponding author.}

\begin{abstract}

Two single channel detection units for thermal neutron detection are investigated in a neutron beam. They consist of two ZnS/${}^6$LiF scintillating layers sandwiching an array of WLS fibers. The pattern of this units can be repeated laterally and vertically in order to build up a one dimensional position sensitive multi-channel detector with the needed sensitive surface and with the required neutron absorption probability. The originality of this work arises from the fact that the WLS fibers are read out with SiPMs instead of the traditionally used PMTs or MaPMTs. The signal processing system is based on a photon counting approach. For SiPMs with a dark count rate as high as 0.7~MHz, a trigger efficiency of 80\% is achieved together with a system background rate lower than $10^{-3}$~Hz and a dead time of 30~$\muup$s. No change of performance is observed for neutron count rates of up to 3.6~kHz.

\end{abstract}


\end{frontmatter}


\section{Introduction}

A one dimensional position sensitive detector for thermal neutrons is currently under development at PSI, in the framework of the upgrade of the engineering diffractometer POLDI \cite{stuhr} installed at the Swiss neutron spalation source (SINQ) at PSI. To allow a simultaneous measurement of the axial and transverse strain components during in-situ deformation measurements, the actual ${}^3$He detector module will be replaced by two oppositely placed scintillation detector modules. The main requirements for the new detector are: a detection efficiency of 65\% at 1.2~$\AA$, a time resolution below 1~$\muup$s (standard deviation), a channel width of about 2.5~mm, a system background rate below $3 \cdot 10^{-3}$~Hz/channel, a sustainable count rate of 4~kHz/channel and a gamma sensitivity below $10^{-6}$.

The worldwide supply shortage of ${}^3$He which started in 2009 stimulated the development of alternatives to ${}^3$He gas detectors for thermal neutron detection \cite{Zeitelhack}. One of this alternatives is based on ZnS/${}^6$LiF or ZnS/${}^{10}$B${}_2$O${}_3$ scintillators combined with wavelength shifting (WLS) fibers for collecting the scintillation light. This technology is now widely used and well established. The two single-crystal time-of-flight diffractometers {\it SENJU} \cite{SENJU,modules_for_SENJU} and iBIX \cite{iBIX} at J-PARC, as well as the engineering diffractometer VULCAN \cite{VULCAN-2,VULCAN-1} and the powder diffractometer POWGEN3 \cite{POWGEN3} in ORNL are examples of instruments currently in operation which are equiped with detectors based on this technology. Up to now, all the detectors of this kind use photomultiplier tubes (PMTs) or multi-anode photomultiplier tubes (MaPMTs) to read out the WLS fibers. The feasibility to use silicon photomultipliers (SiPMs) has not been proved yet. Nevertheless, the use of SiPMs for the new POLDI detector is very attractive because of their high packing fraction and their insensitivity to magnetic fields, two important characterisitcs in this project since the space in the POLDI experimental area is very limited and tests of samples in high magnetic fields are foreseen. 

Two variants of a single channel detection unit which can constitute an elementary building block for a one dimensional position sensitive multi-channel detector have been investigated in a neutron beam. The detection units consist of ZnS/${}^6$LiF scintillating layers with embedded WLS fibers read out with a silicon photomultiplier (SiPM). In this paper, we present the design of the detection units, the signal processing system and the measurements of the probability of neutron capture, the light yield distribution for the capture of neutrons, the trigger efficiency, the system background rate, the count rate capability.

\section{Material and method}

\subsection{The detection units}

\begin{figure}[h] 
\centering
\includegraphics[width=21pc]{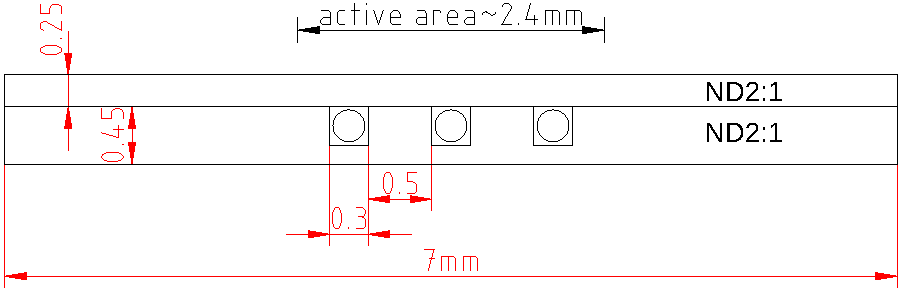}
\includegraphics[width=21pc]{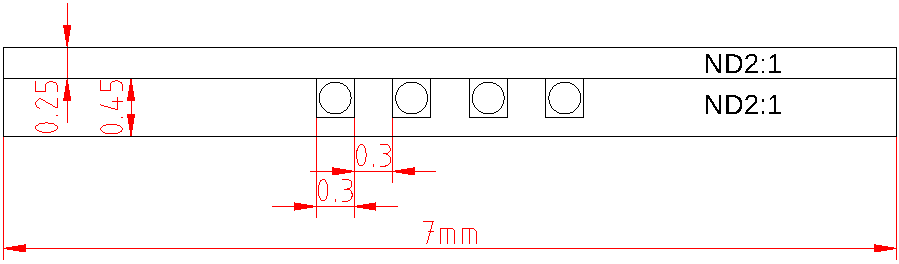}
\caption{Cross-section of the detection units with three fibers (top) and four fibers (bottom).}
\label{fig:schematic_sandwich}
\end{figure}

Figure \ref{fig:schematic_sandwich} shows the schematic of the two detection units which were tested in the present work. They consist of two ZnS/${}^6$LiF scintillating layers (ND2:1 from Applied Scintillation Technologies \cite{AST}) sandwiching an array of WLS fibers (Y11(200)M from Kuraray \cite{kuraray}) with a diameter of 250~$\muup$m. The fiber pitch is 0.6~mm and 0.8~mm for the 4-fiber and 3-fiber detection units, respectively. In the 450~$\muup$m thick layer (bottom layer), three or four grooves of 300~$\muup$m width and 300~$\muup$m depth are machined. The fibers are glued into the grooves with an optical epoxy (EJ500 from Eljen technology \cite{ELJEN}) and the top scintillating layer 250~$\muup$m thick is glued with the same optical epoxy.

\begin{figure}[h]
\begin{center}
\begin{minipage}{10pc}
\begin{center}
\includegraphics[width=9pc]{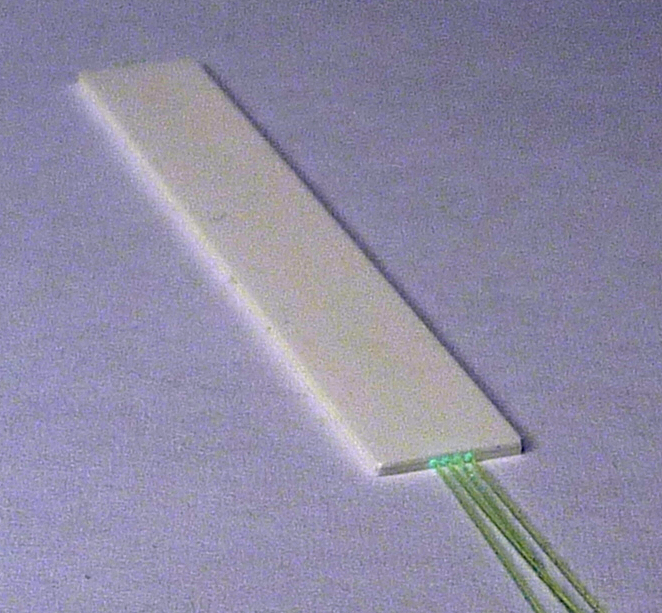}
\caption{The 4-fiber detection unit before assembling with a fiber holder.}
\label{fig:sandwich_before_gluing}
\end{center}
\end{minipage}
\hspace{0.pc}
\begin{minipage}{10pc}
\begin{center}
\includegraphics[width=4pc]{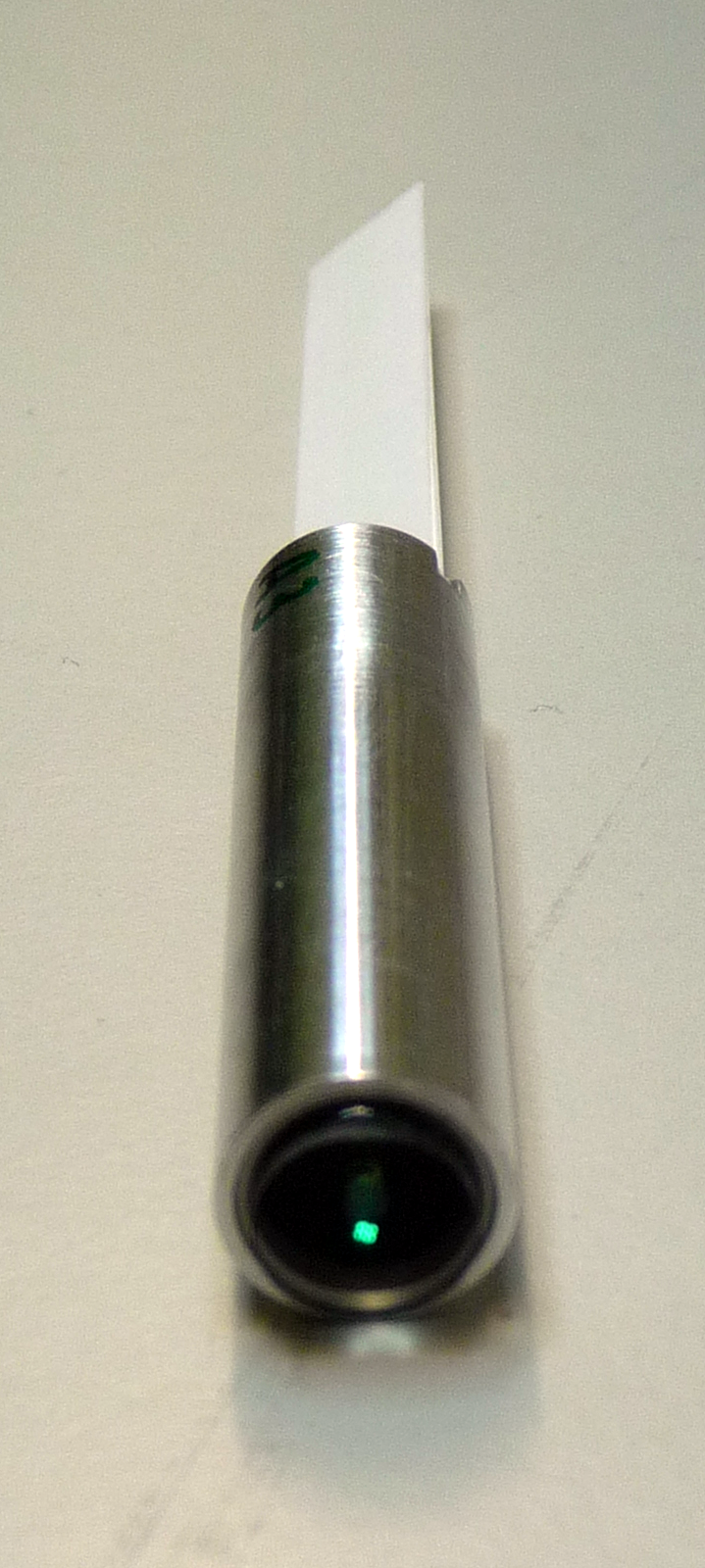}
\caption{A detection unit ready to be connected to a SiPM.}
\label{fig:sandwich_after_gluing}
\end{center}
\end{minipage}
\end{center}
\end{figure}

On one side, the fibers are cut along the edge of the sandwich and polished. Then, an aluminum foil acting as a mirror is glued to improve the light yield on the other side of the fibers where the photodetector is connected. On this side, the fibers are glued together into the hole of a plexiglas holder and polished. Fig. \ref{fig:sandwich_before_gluing} and \ref{fig:sandwich_after_gluing} show a picture of the sandwich before and after assembling with a fiber holder.

The effective channel width is defined by the area covered by the fibers. It is about 2.4~mm. The aditionnal space to the total width of 7~mm is used for handling purposes only. The length of the structure is 50~mm.

\begin{figure}[h] 
\centering
\includegraphics[width=21pc]{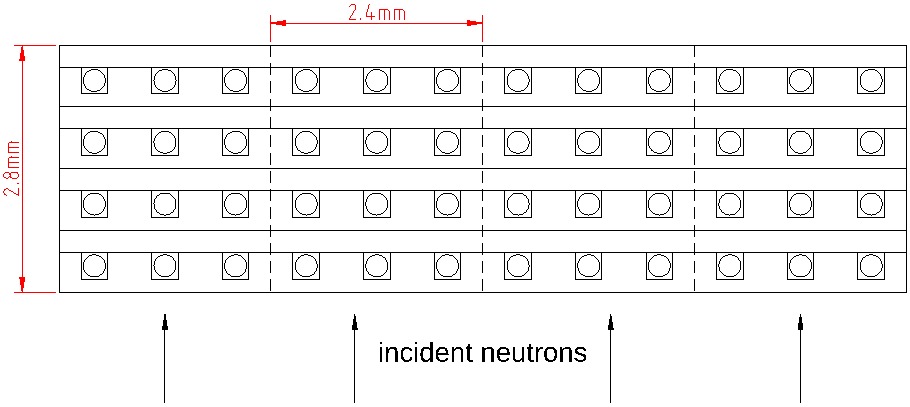}
\caption{Cross-sectional view of a possible multi-channel detector repeating latterally and vertically the pattern of the single channel detection unit with three fibers. The channels are delimited by the dashed lines. Each channel contains 12 WLS fibers.}
\label{fig:multichannel}
\end{figure}

In a later multi-channel detector, this 2.4~mm wide groove/WLS fiber pattern without the additional handling space will be repeated to cover the needed sensitive area. In order to obtain the required neutron absorption probability, several such sandwiches will be stacked together (Fig. \ref{fig:multichannel}). A 4-sandwich detector would have a thickness of 2.8~mm and its intrinsic time resolution would be below 0.4~$\muup$s for all neutron wavelengths between 1~$\AA$ and 6~$\AA$. By the intrinsic time resolution, we understand the standard deviation of the neutron travel time from the moment it enters the detector to the moment it is absorbed in it.

The ${}^6$Li concentration in the scintillating layers has been determined with a neutron transmission measurement. It amounts to $1.4 \cdot 10^{22}$~atoms/cm${}^3$. The hydrogen concentration amounts to about $2.4 \cdot 10^{22}$~atoms/cm${}^3$ and $4.8 \cdot 10^{22}$~atoms/cm${}^3$ in the scintillating layers and in the WLS fibers, respectively.

The SiPM used in this test is a $1\times1$~mm${}^2$ S12571-025C MPPC from Hamamatsu \cite{hamamatsu}. It is operated at room temperature, at the recommended overvoltage of 3.5~V. In this condition, the SiPM has a PDE of 35\% at 480~nm (the emission peak of the WLS fibers), a crosstalk probability of 22\% \cite{hamamatsu}, a dark count rate of 70~kHz and an afterpulse probability below 3\% \cite{SiPM_improvement}.

\subsection{The signal processing system}

The signal processing system is based on a photon counting approach. Two reasons make this approach feasible. First, the ZnS/$^6$LiF scintillator is relatively slow and consequently, the photons to detect are sufficiently spaced out in time to be counted individually. Secondly, the SiPM have an excellent photon counting capability.

\begin{figure}[h] 
\centering
\includegraphics[width=21pc]{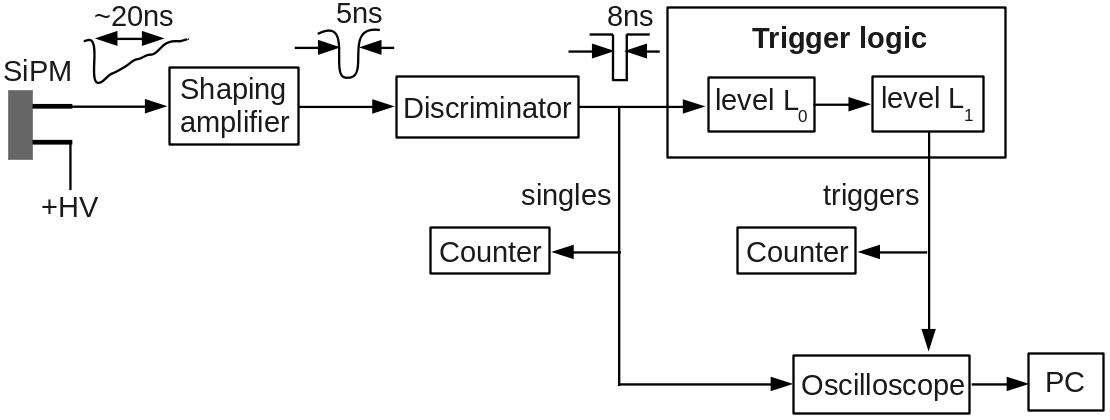}
\caption{Block diagram of the signal processing system.}
\label{fig:SPS_schematic}
\end{figure}

Fig. \ref{fig:SPS_schematic} shows the block diagram of the signal processing system. The SiPM signal is amplified and shaped by a wide band-width amplifier. The output of the amplifier is fed into a fast leading edge discriminator with a threshold set at 0.5 photoelectron. The discriminator is set in burst guard mode so that its output signal stays high as long as the input is over the threshold. The duration of the output for a single photoelectron is 8~ns.

The trigger logic is composed of two levels L${}_0$ and L${}_1$. The principle of the level L${}_0$ is the following. Each count from the discriminator opens a time window W${}_1$ of duration T${}_1$. If the next count arrives during W${}_1$, a second time window W${}_2$ of duration T${}_2$ is opended. The level L${}_0$ triggers if the next count arrives during W${}_2$. Then a third time window W${}_3$ of duration T${}_3$ equal to 2$\muup$s is opended. The level L${}_1$ triggers if the number of counts during W${}_3$ is equal or higher than a threshold value n${}_3$. Table \ref{trigger_settings} shows the two trigger settings used in this work. The low threshold trigger used for the measurements of the photoelectron number distribution at low count rates requires in total a minimum of 7 photoelectrons within a time window of 2.8~$\muup$s and the high threshold trigger used for the measurements at high count rates requires in total a minimum of 23 photoelectrons within a time window of 2.3~$\muup$s.

In order to prevent multiple triggers on the same neutron event, the trigger logic needs to be blocked after each trigger during some time. The minimum blocking time depends on the trigger setting. At low and high threshold, the blocking time is set at 200~$\muup$s and 30~$\muup$s, respectively.

The capture of a neutron within this blocking time leads to the risk to get a late trigger related to this capture after the trigger logic is released. For this reason, the measurements with the long blocking time of 200~$\muup$s are performed at a small rate of about 20~Hz which makes this process negligible.

\begin{table}[h]
\caption{Trigger settings used in this work.}
\begin{center}
\begin{tabular}{ccccc}
\hline
Trigger setting & T${}_1$ (ns) & T${}_2$ (ns) & T${}_3$ (ns) & n${}_3$ in W${}_3$\\
\hline
low threshold & 300 & 500 & 2000 & 4\\
high threshold & 100 & 200 & 2000 & 20\\
\hline
\end{tabular}
\label{trigger_settings}
\end{center}
\end{table}

When a trigger is generated, the digital signal after the discriminator is sampled during 80~$\muup$s at a sampling rate of 250~MHz with a digital scope 3206MSO from Pico Technology \cite{pico_technology}. Then, the sampled sequence is sent to a PC. The maximum acquisition rate of the oscilloscope is about 30~Hz. The oscilloscope is set to acquire sampled sequences starting 3.2~$\muup$s before the arrival of the trigger. In this way, the first count of a bunch of counts producing a trigger is recorded in the sampled sequence.

A sampled sequence consists of 20000 samples with values equal to 0 or 1. A single photon generates a group of 2 or 3 consecutive samples with a value equal to 1. During the offline analysis of the data, the 20000 samples are read from the first one to the last one. The occurence of a sample with a value equal to 1 is seen as a photon or a dark count. Then, the next 2 samples are skipped. In other words, an artificial dead time of 12~ns is introduced in the photon counting process.

For each trigger accepted by the oscilloscope, the sampled sequence provides the time stamp of each photoelectron and dark count measured between the instant of the neutron capture and 80~$\muup$s later. This information allows to determine the light yield distribution for differrent time windows. In this work, the following two windows are considered: [0;2$\muup$s] and [2$\muup$s;12$\muup$s].

The time origin of the time scale used to define this windows should correspond approximately to the instant of the neutron capture. In order to determine the time origin, the first 3.2~$\muup$s of the sampled sequence are divided into 32 consecutive time slots $S_{i}$ ($i=1,... ,32$) of 100~ns each. Then, the slot $S_{i}$ with the maximum number of counts is identified. If $i\ne1$, the first count occuring in $S_{i-1}$ gives the time origin. If $i=1$ or if there is no count in $S_{i-1}$, the time origin is given by the first count in the slot $S_{i}$.

It must be stressed that, in comparison to a signal processing system based on a slow shaping amplifier followed by a discriminator, a photon counting based signal processing system offers a significant advantage because it allows to get rid of the SiPM crosstalk. Indeed, when several pixels of a SiPM are fired at the same time, only one count is recorded.

Let's consider a SiPM having the same dark count rate of 70~kHz and the same crosstalk probability of 22~\% as the SiPM used in this work. To get an insight into the impact of the artificial crosstalk suppression on the system background rate, the probability distribution of the number $n$ of fired cells measured during 2~$\muup$s time intervals has been determined without and with crosstalk suppression. In the latter case, the distribution is easy to calculate since it simply follows the Poisson statistics. Without crosstalk suppression, the distribution is obtained with a toy Monte Carlo in which the interpixel crosstalk is implemented as described in Ref. \cite{balagura}.

\begin{figure}[h] 
\centering
\includegraphics[width=22pc]{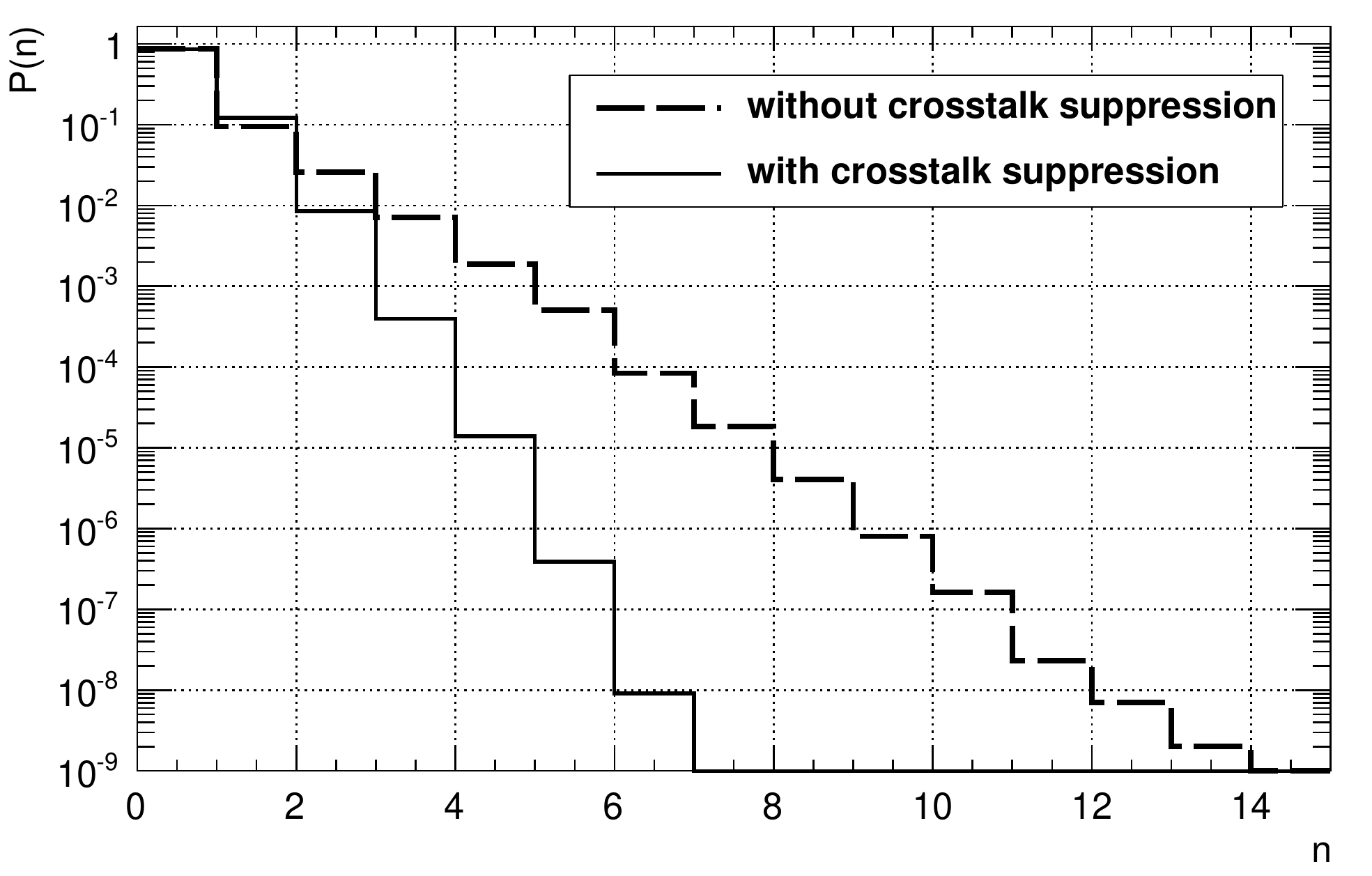}
\caption{Probability distribution of the number $n$ of fired cells measured during 2~$\muup$s time intervals, without and with crosstalk suppression.}
\label{fig:suppression_crosstalk}
\end{figure}

 Fig. \ref{fig:suppression_crosstalk} shows that the probability distribution has a much smaller tail when the crosstalk is suppressed. Consequently, for a given threshold, the crosstalk suppression allows to reduce significantly the system background rate. For example, the probability to measure a system background event with more than five fired cells is about four orders of magnitude lower when the crosstalk is suppressed.

\subsection{Experimental set-up}

The measurements are performed in the ORION beamline at the Swiss neutron spallation source (SINQ) at the Paul Scherrer Institut (PSI). The neutron beam is monochromatic and the wavelength of the neutrons is equal to $2.2~\AA$. A beam monitor is used to control the stability of the beam intensity during the measurements.

A neutron attenuating mask is mounted on the detection units to avoid the capture of neutrons outside the fully active area of the detection units. The mask is made of a 3.2~mm thick neutron shielding material (JC238 Flex-Boron from JCS \cite{JCS}). This material provides an attenuation factor of 259 for thermal neutrons. The window of the mask has an area of 40~mm $\times$ 1.5~mm which is centered on the active area of the detection units.

The rate of neutrons reaching the detection units is set with a variable slit width. The surface of the detection units is oriented perpendicularly to the neutron beam and the bottom layer of the detection units is facing the beam.

\section{Results and discussion}

\subsection{Distribution of the number of photoelectrons}

\begin{figure}[h!]
\centering
\includegraphics[width=21pc]{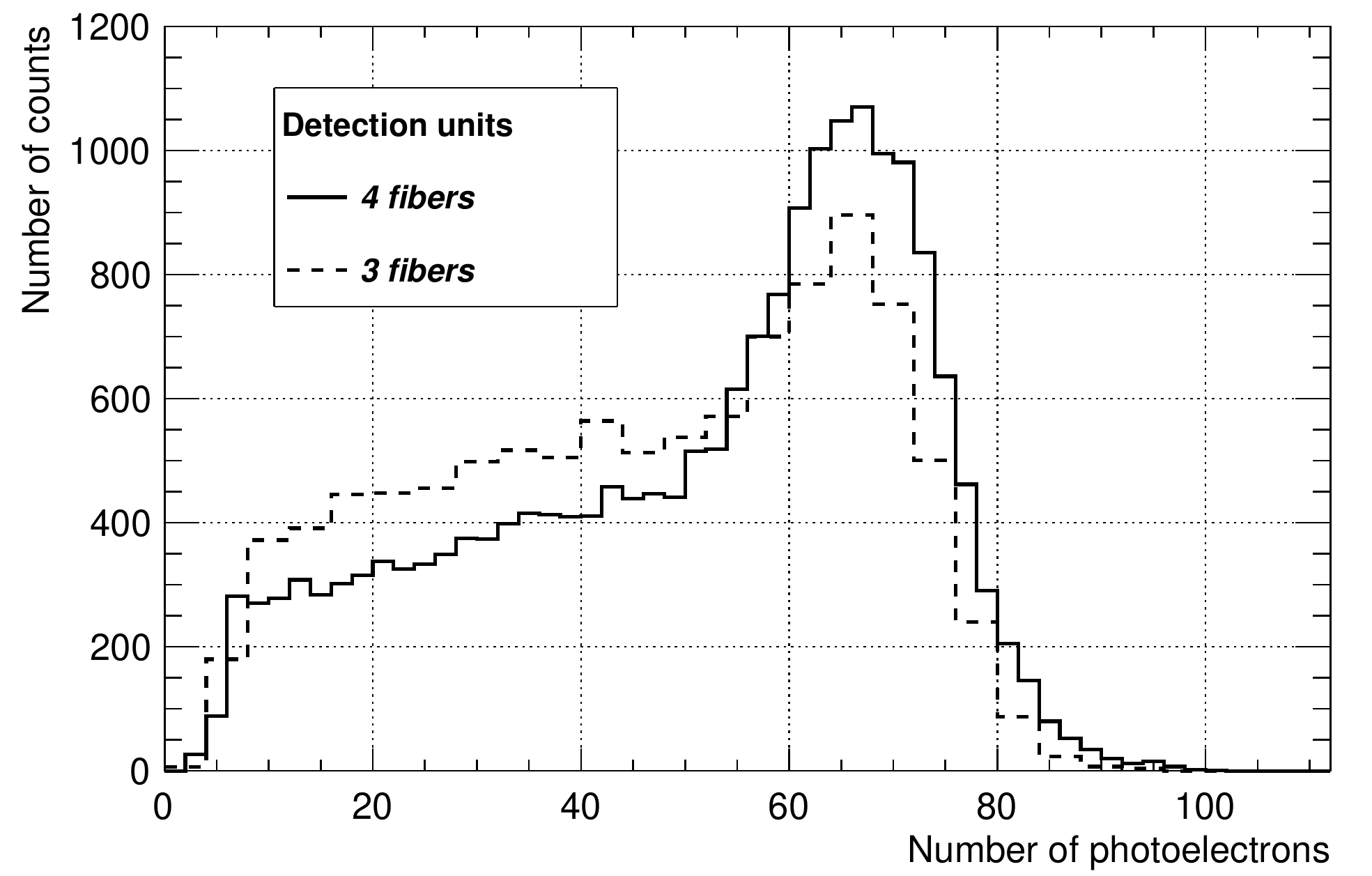}
\caption{Photoelectron number distributions for the [0;2$\muup$s] time window.}
\label{fig:Npe_spectrum_0_2}
\end{figure}

Fig. \ref{fig:Npe_spectrum_0_2} shows the distribution of the number of photoelectrons measured with the two detection units, for the [0;2$\muup$s] time window. Not surprisingly, the average number of photoelectrons is slightly higher with 4 fibers than with 3 fibers. The peak of each distribution is the consequence of the saturation of the photoelectron counting process which has a non-extendable dead time of 12~ns.

\begin{figure}[h!]
\centering
\includegraphics[width=21pc]{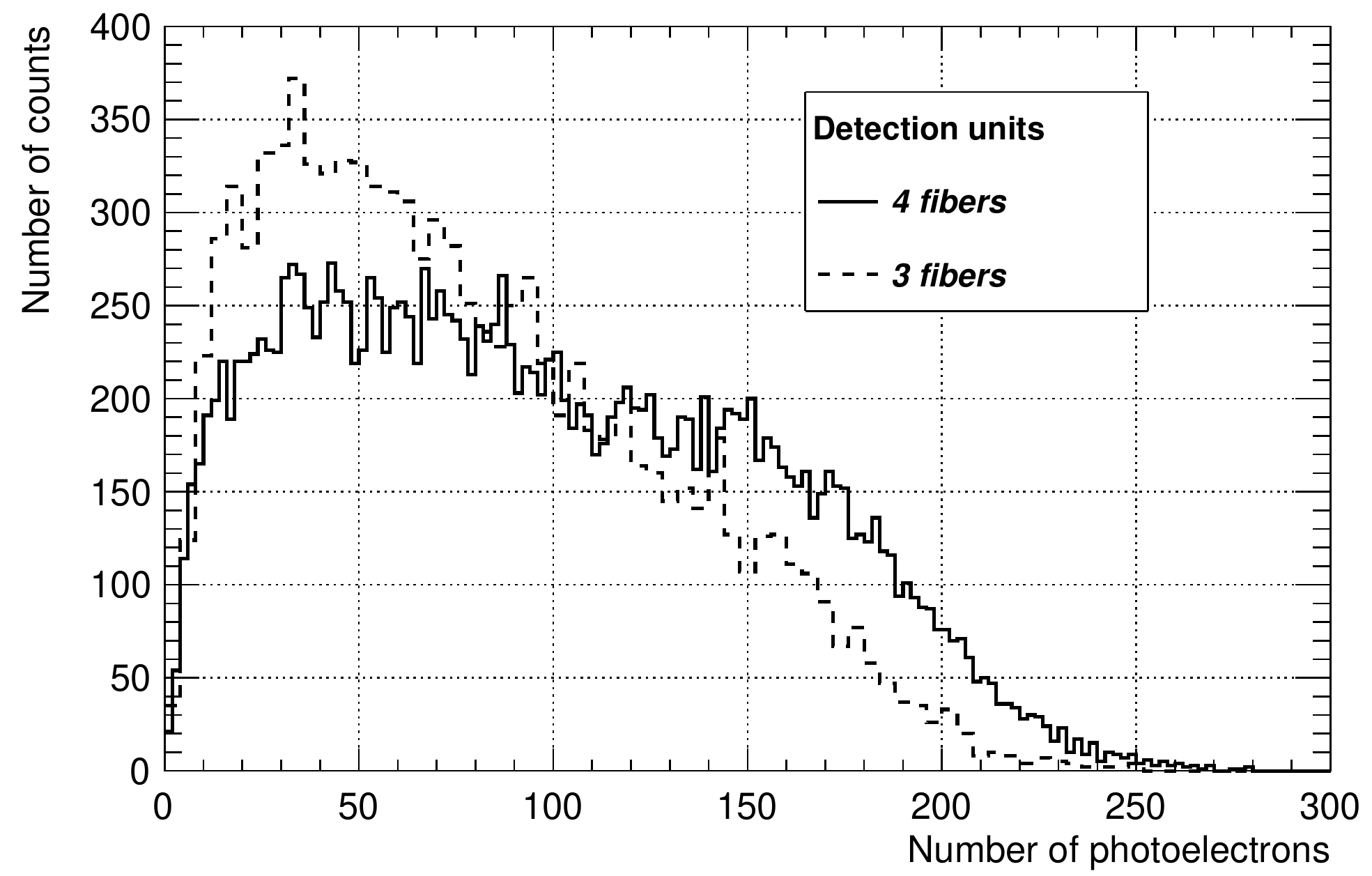}
\caption{Photoelectron number distributions for the [2$\muup$s;12$\muup$s] time window.}
\label{fig:Npe_spectrum_2_12}
\end{figure}

Fig. \ref{fig:Npe_spectrum_2_12} shows the photoelectron number distribution for the [2$\muup$s;12$\muup$s] time window. In this time window, the photoelectron density is much smaller than in the [0;2$\muup$s] time window and the counting losses are sufficiently small to allow the correction of the photoelectron number distribution for the dead time of the counting process.

\begin{figure}[h!]
\centering
\includegraphics[width=21pc]{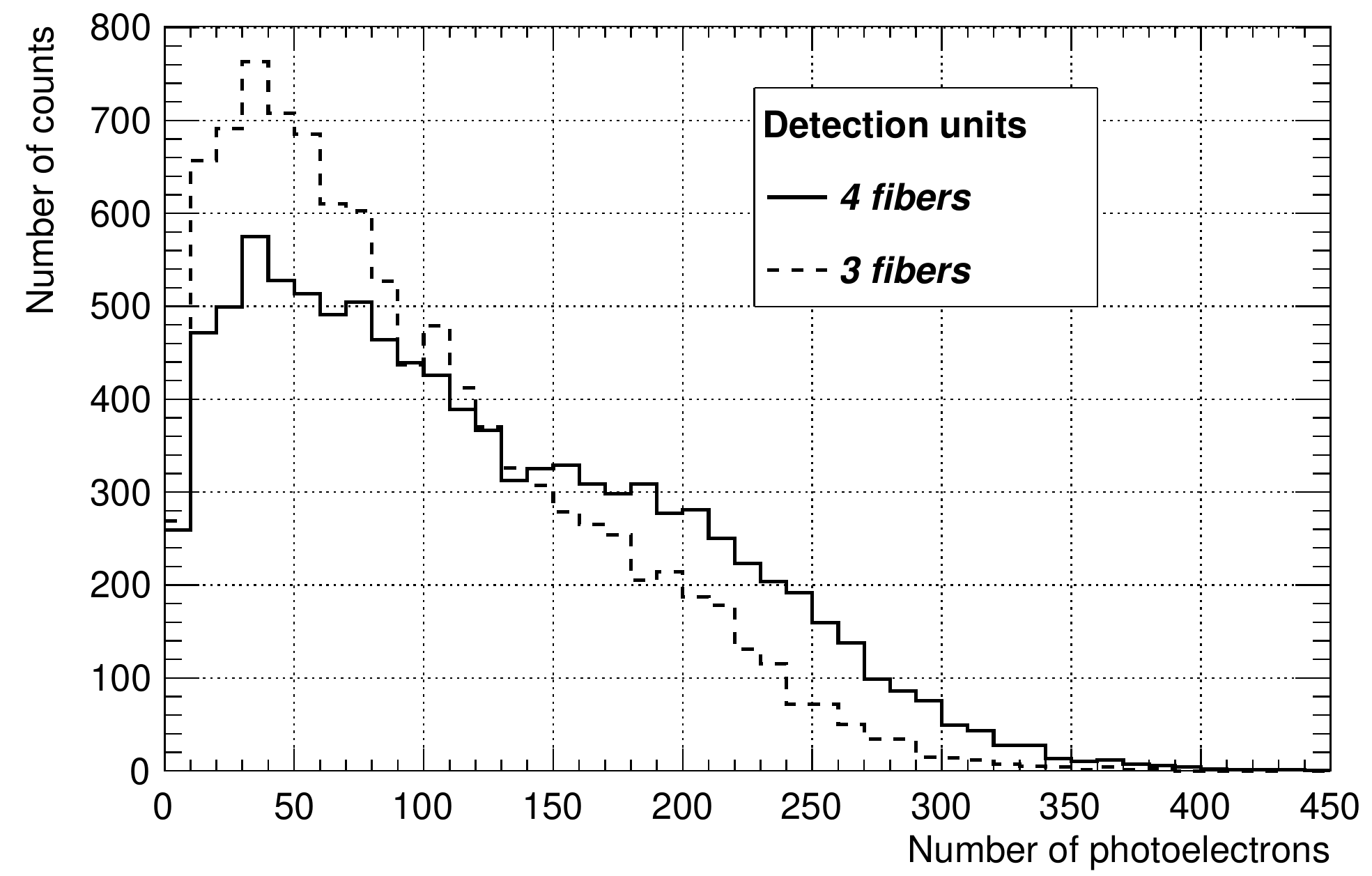}
\caption{Photoelectron number distributions for the [2$\muup$s;12$\muup$s] time window after correction for the dead time of the photoelectron counting process.}
\label{fig:true_Npe_spectrum}
\end{figure}

The correction considers, event by event, the histogram of the photoelectron time distribution measured in the [2$\muup$s;12$\muup$s] time window. For each bin of the histogram, the dead time corrected photoelectron number $n$ is calculated with the equation
\begin{equation}
n = \frac{m}{1-m\tau / \Delta t} \label{eq:dead_time_correction}
\end{equation}
where $m$ is the measured photoelectron number, $\tau$ is the dead time and $\Delta t$ is the bin width. The sum of all this corrected values provides an estimate of the true number of photoelectrons in the [2$\muup$s;12$\muup$s] time window. Fig. \ref{fig:true_Npe_spectrum} shows the true photoelectron number distributions for the [2$\muup$s;12$\muup$s] time window. The shape of these distributions correspond to the shape of the light yield distributions.

\subsection{Neutron detection efficiency}

The detection efficiency is the product of the absorption probability and the trigger efficiency of the signal processing system. The absorption probability is simply related to the number of sandwiches stacked together. For a 4-sandwich detector (Fig. \ref{fig:multichannel}), the probability of direct capture of primary neutrons at a wavelength of $1.2~\AA$ would amount to 70\%. Due to the rather high hydrogen concentration in the detection volume, the scattering probability would amount to about 20\%. However, 60\% of this scattered neutrons would be later on absorbed in the same detection channel. So, at a wavelength of $1.2~\AA$, the neutron capture probability is estimated to be 82\%. The estimation of the probability of direct capture is based on an analytical calculation and the estimation of the fraction of scattered neutrons which are later on captured in the same detection channel is based on a toy Monte Carlo simulation.

The determination of the trigger efficiency is based on the extrapolation of the histogram of the photoelectron number distribution measured with the [0;2$\muup$s] time window. From 0 to 6~photoelectrons, the bin content is set to the average amplitude of the histogram between 7 and 12~photoelectrons. The trigger efficiency $\epsilon_{trigger}$ is then given by the equation
\begin{equation}
\epsilon_{trigger} = 1 - \frac{\int_{0}^{PE_{threshold}} h}{N_{events}} \label{eq:trigger_efficiency}
\end{equation}
where $PE_{threshold}$ is the threshold value on the number of photoelectrons, $h$ is the extrapolated histogram of the photoelectron number distribution and $N_{events}$ is the number of events in $h$. 

\begin{figure}[h!]
\centering
\includegraphics[width=21pc]{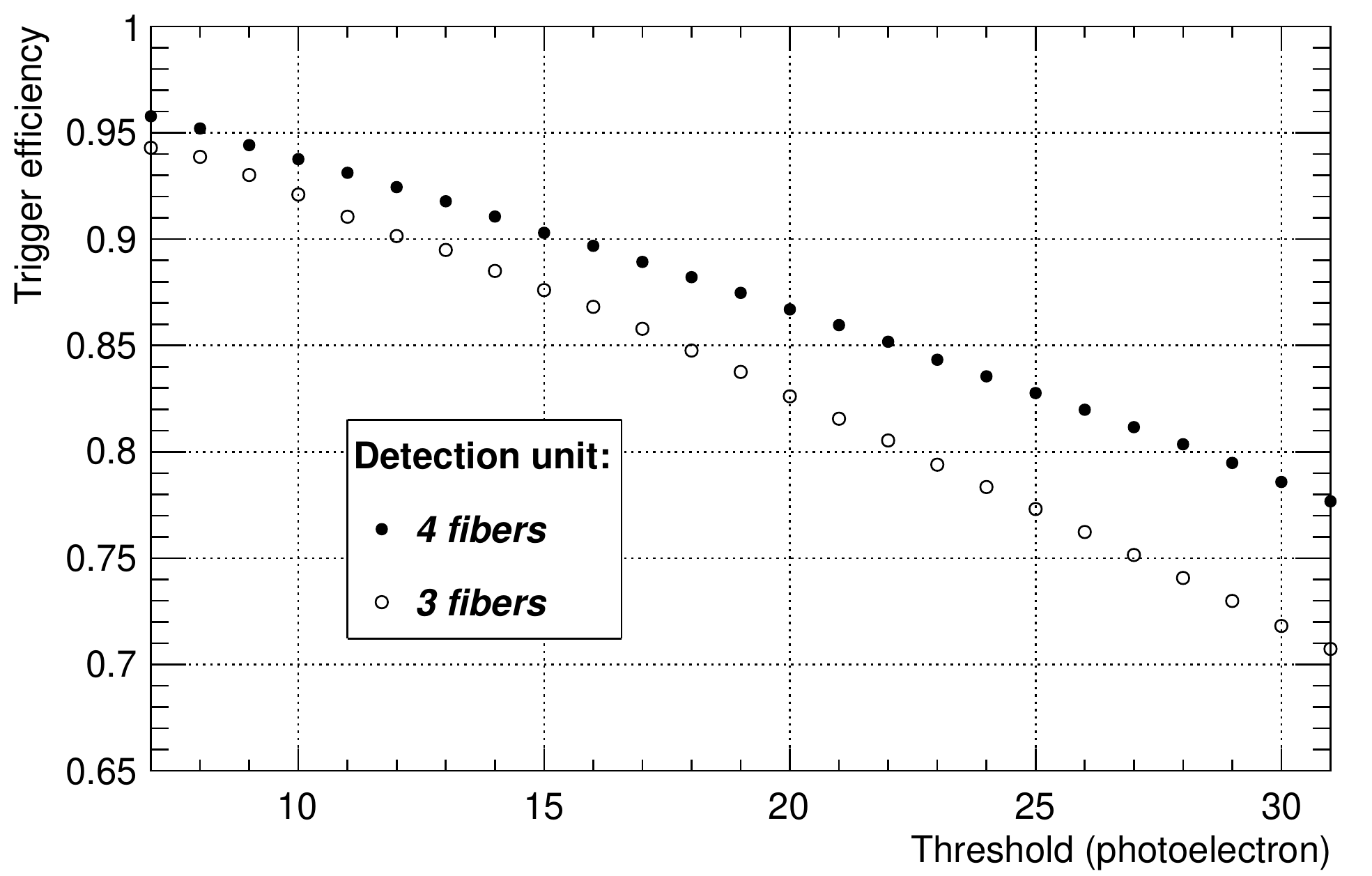}
\caption{Trigger efficiency as a function of the threshold on the photoelectron number.}
\label{fig:trigger_efficiency_vs_threshold}
\end{figure}

Fig. \ref{fig:trigger_efficiency_vs_threshold} shows the trigger efficiency as a function of the threshold on the number of photoelectrons for the two detection units. The detection unit with four fibers has a higher trigger efficiency. This is the consequence of the more efficient light collection.

\begin{figure}[h!]
\centering
\includegraphics[width=21pc]{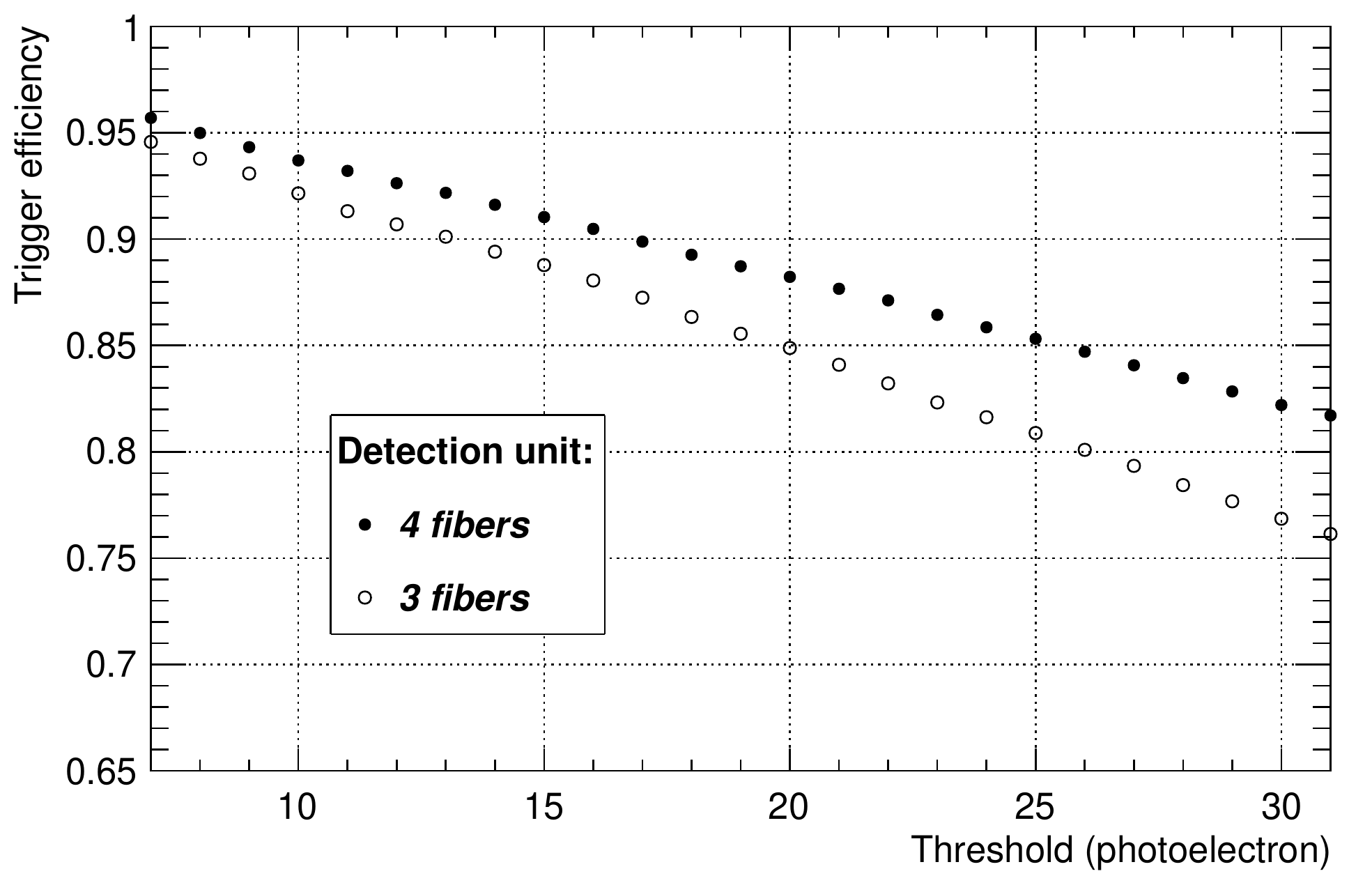}
\caption{Expected trigger efficiency as a function of the threshold on the photoelectron number for the dead time in the photoelectron counting process equal to 5~ns instead of 12~ns.}
\label{fig:trigger_efficiency_vs_threshold_dead_time_corrected}
\end{figure}

Reducing the dead time of the photoelectron counting process would slightly increase the trigger efficiency. Fig. \ref{fig:trigger_efficiency_vs_threshold_dead_time_corrected} shows the expected trigger efficiency as a function of the threshold on the number of photoelectrons, for a dead time of 5~ns instead of 12~ns. The data are obtained by using the equation \ref{eq:trigger_efficiency} with the extrapolation of the dead time corrected photoelectron distribution. Up to 30 photoelectrons, the counting losses are sufficiently small to allow the correction for the dead time.

\subsection{System background rate}

\begin{figure}[h!]
\centering
\includegraphics[width=21pc]{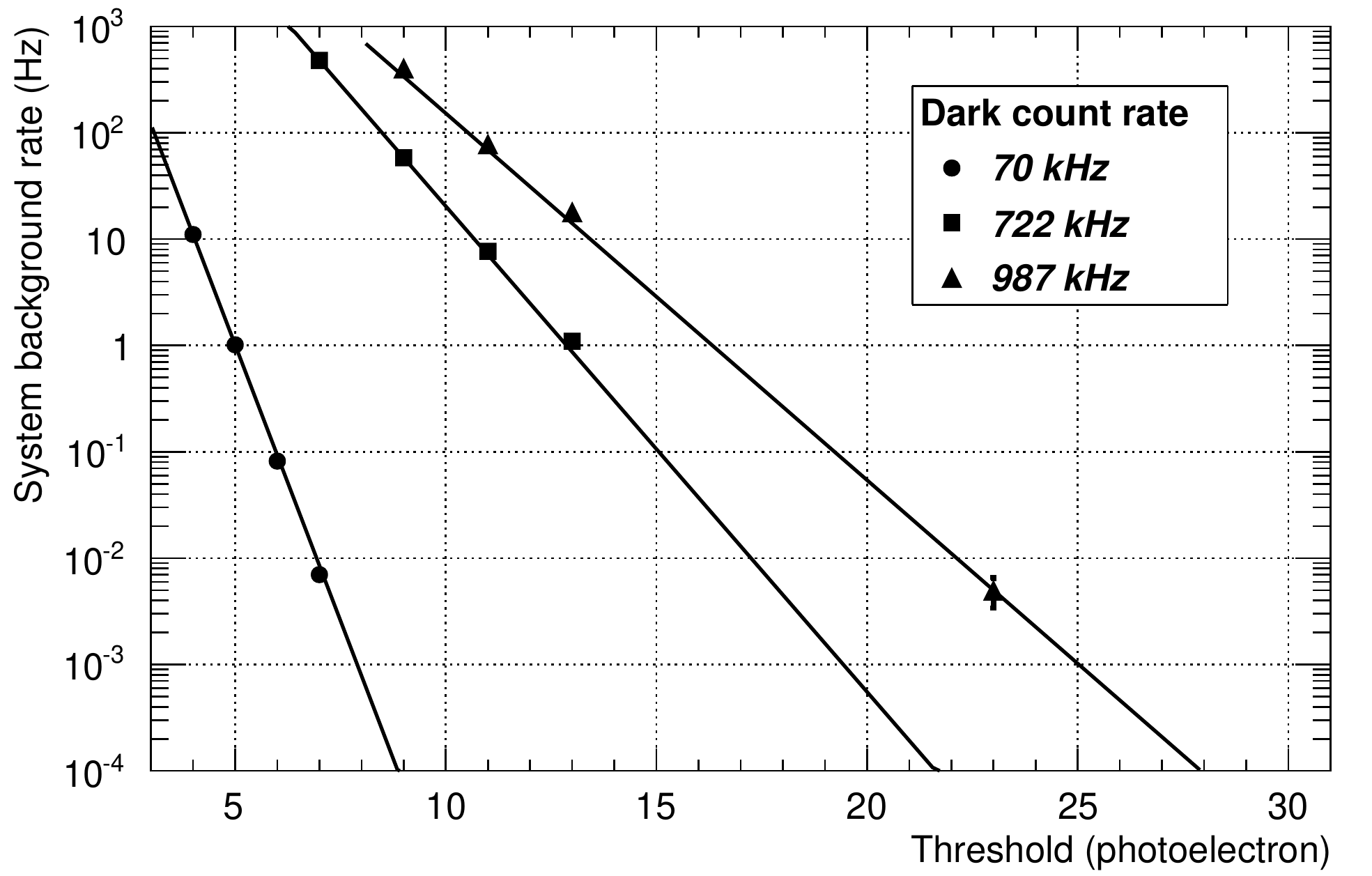}
\caption{System background rate as a function of the threshold on the photoelectron number at different SiPM dark count rates. The system background rate is fitted with the function $\it{f(x) = \alpha \cdot e^{-\beta \cdot x}}$.}
\label{fig:noise_vs_threshold}
\end{figure}

The nominal SiPM dark count rate of 70~kHz at 25$\degree$C is relatively low. However, during operation of the detection system, the dark count rate might undergo short-term variations caused by variations of the SiPM temperature, and in long-term it might increase due to radiation damage \cite{radiation_damage}. In order to increase artificially the dark count rate of our non-irradiated SiPM, it is illuminated with a LED driven at a DC bias voltage. Fig. \ref{fig:noise_vs_threshold} shows the system background rate as a function of the threshold on the photoelectron number at different SiPM dark count rates. At a dark count rate of 722~kHz and with a threshold set at 23 photoelectrons, the number of system background counts measured over 4000~s is equal to 0, indicating that the system background rate is well below $10^{-3}$~Hz.

\subsection{Count rate capability}

The dead time of the detector is equal to the width of the protection window which disables the trigger logic after each trigger in order to prevent multiple triggers on the same neutron event due to scintillator afterglow.

The required width of the protection window strongly depends on the threshold value. For a threshold set at 7~photoelectrons, a protection window of 200~$\muup$s is required while for a threshold set at 23~photoelectrons, a protection window of 30~$\muup$s is sufficient.

In order to optimize the width of the protection window at a threshold value equal to 23~photoelectrons, the trigger rate is measured at a constant neutron intensity (Table \ref{tab:double_triggers}). No statistically significant variation of the trigger rate is measured by reducing the width of the protection window from 50~$\muup$s to 30~$\muup$s while an increase of the trigger rate of about 1\% and 14\% is measured by reducing the protection window from 50~$\muup$s to 20~$\muup$s and from 50~$\muup$s to 10~$\muup$s, respectively.

\begin{table}[h]
\caption{Rate of triggers as a function of the width of the protection window at a threshold value equal to 23~photoelectrons.}
\begin{center}
\begin{tabular}{ccccc}
\hline
Width of the protection window ($\muup$s) & Trigger rate (Hz) \\
\hline
50 & $21.30\pm0.26$ \\
30 & $21.35\pm0.26$ \\
20 & $21.57\pm0.26$ \\
10 & $24.10\pm0.26$ \\
\hline
\end{tabular}
\label{tab:double_triggers}
\end{center}
\end{table}

\begin{figure}[h!]
\centering
\includegraphics[width=21pc]{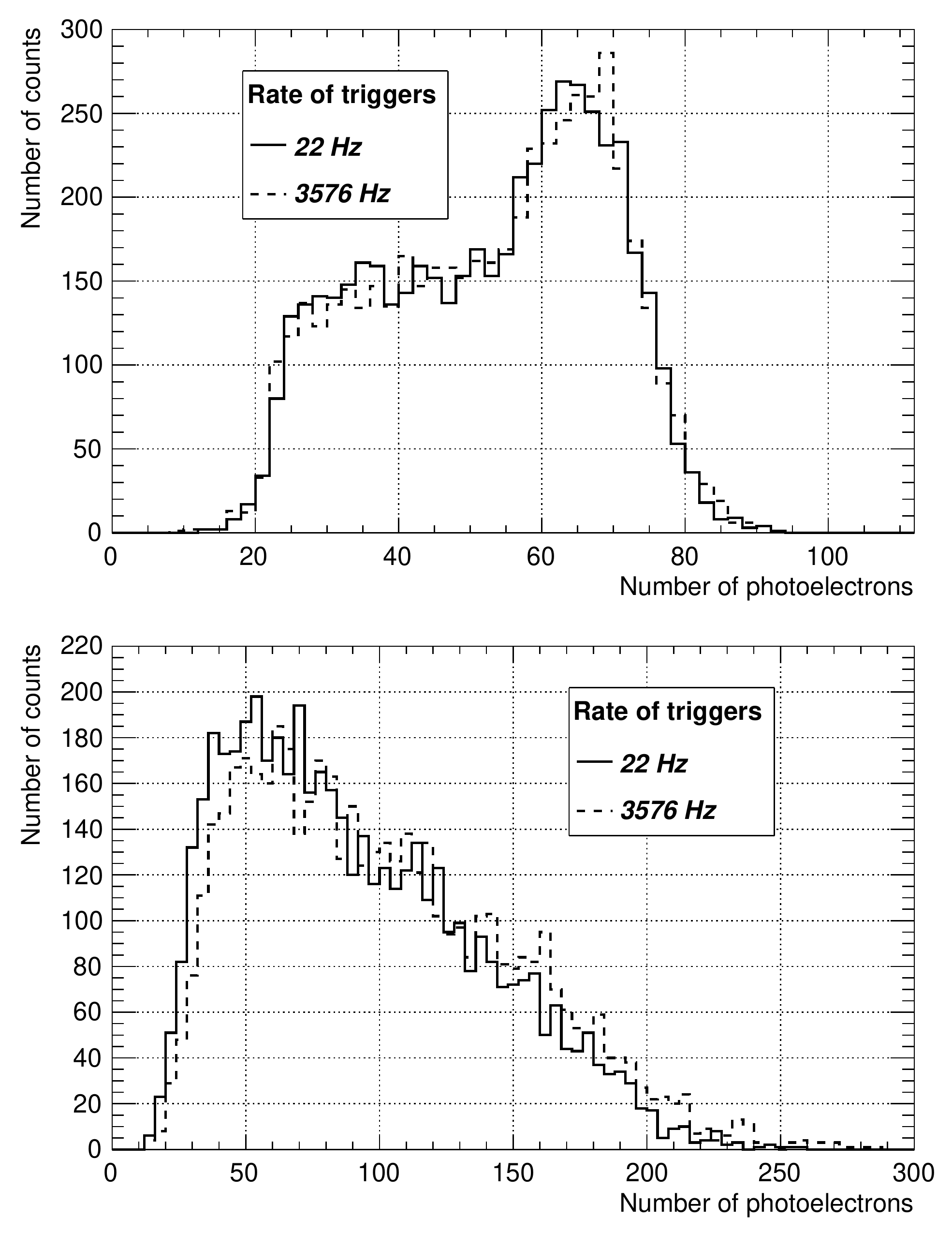}
\caption{Photoelectron number distributions measured with the 3-fiber detection unit at low and high count rates of neutrons for two different time windows: [0;2$\muup$s] (top), [2$\muup$s;12$\muup$s] (bottom).}
\label{fig:Npe_spectrum_high_low_rate}
\end{figure}

Fig. \ref{fig:Npe_spectrum_high_low_rate} shows the photoelectron number distribution measured with the 3-fiber detection unit, at low (21~Hz) and high (3576~Hz) count rates of neutrons for the two time windows [0$\muup$s;2$\muup$s] and [2$\muup$s;12$\muup$s]. The distributions are almost the same at low and high neutron rates. However, at high count rate, one can notice a little shift to the right of the whole distribution obtained with the [2$\muup$s;12$\muup$s] time window. This shift corresponds to the integration of the background counts produced by the scintillator afterglow.

Knowing the rate of singles $\it{N_{singles}}$ (1567~kHz), the SiPM dark count rate $\it{N_{SiPM}}$ (70~kHz) and the rate of captured neutrons $\it{N_{neutrons}}$ (5007~Hz) obtained after correction of the measured neutron rate for the dead time of the detection unit and for the 80\% trigger efficiency, equation \ref{eq:PE_number_per_neutron} allows to estimate the average number of photoelectrons $\it{n_{PE}}$ produced by a neutron. The calculation gives a value of about 300~photoelectrons per captured neutron.

\begin{equation}
n_{PE} = \frac{N_{singles} - N_{SiPM}}{N_{neutrons}} \label{eq:PE_number_per_neutron}
\end{equation}

The information concerning the decay curve of the scintillator \cite{dubna} shows that 23\% of the scintillating light is produced after the protection window of 30~$\muup$s. Consequently, in average 70~photoelectrons are produced after the protection window, resulting in a significant photoelectron background rate which increases proportionally with the rate of neutrons. Thus, during a measurement performed at rate of captured neutrons of 5~kHz, the photoelectron background rate due to the scintillator afterglow is about 340~kHz, which is already 5 times higher than the SiPM dark count rate. 

\section{Conclusion}

Two detection units for thermal neutrons combining ZnS(Ag)/${}^6$LiF scintillating layers and WLS fibers read out with a SiPM have been evaluated in a neutron beam.

By repeating lateraly the pattern of this detection units, it is possible to build a one dimensional position sensitive multi-channel detector with the needed sensitive area. In order to get the required neutron detection efficiency for the new detector of the POLDI diffractometer (65\% at 1.2~$\AA$), the pattern with 3 fibers as well as the pattern with 4 fibers should be repeated vertically at least 4 times, leading to a number of fibers per channel equal to 12 and 16, respectively. While 12 fibers can fit into a 1.00~mm diameter hole which matches almost perfectly the $1\times1$~mm${}^2$ sensitive area of the SiPM, 16 fibers can only fit into a 1.15~mm diameter hole which goes over the edge of the sensitive surface of the SiPM. The use of a SiPM with a larger sensitive surface is not desirable since in that case, the SiPM would have a higher initial dark count rate and it would accumulate a higher radiation dose. Consequently, the 3-fiber pattern is prefered, even if it provides a slightly lower light collection efficiency which results in a sligthly lower trigger efficiency in comparison to the 4-fiber pattern.

For SiPMs with a dark count rate as high as 0.7~MHz, the 3-fiber detection unit provides a trigger efficiency of 80\% together with a system background rate below $10^{-3}$~Hz, a dead time of 30~$\muup$s and a negligible rate of multiple triggers. No change of performance is observed for neutron count rates of up to 3.6~kHz.

Apart from the gamma sensitivity which has not been measured yet, the obtained results indicate that a multi-channel detector based on the 3-fiber pattern could satistfy without compromise all the requirements for the new POLDI detector.

When the dead time of the 3-fiber detection unit is set to 30~$\muup$s or less, the level of the trigger threshold is conditioned by the maximum acceptable rate of multiple triggers and not by the maximum acceptable system background rate which is related to the SiPM dark count rate. So, in this case the limiting factor for the trigger efficiency is not the SiPM but the scintillator itself due to its afterglow.

SiPMs offer many advantages over PMTs and MaPMTs. They are significantly less expensive, they have a high packing fraction, they are available with small active areas (e.g. $1\times1$~mm${}^2$) and they are insensitive to magnetic fields. Thus, the use of SiPMs turns out to be a significant technological breakthrough which opens new possibilities in designing ZnS/${}^6$LiF scintillation neutron detectors.



\section*{Acknowledgements}

This work was supported by the Swiss National Science Foundation under Grant No.~206021-139106/1.

\end{document}